\documentclass[11pt]{article}
\usepackage{graphicx}
\usepackage{url}
\usepackage{longtable}
\usepackage{amsfonts}
\usepackage{amsmath,amsthm,amssymb}

\def\directunion{\hbox{$\bigcirc$ \hskip - 11.3 pt \raise 0.1pt
\hbox{$\scriptstyle \vee$}}\ }
\newtheorem{theorem}{Theorem}
\newtheorem{definition}{Definition}

\newtheorem{proposition}{Proposition}

\begin{document}

\title{Quantum Axiomatics: Topological and Classical Properties \\ of State Property Systems}
\author{Diederik Aerts\thanks{ Centre Leo Apostel and Department of Mathematics, Vrije Universiteit Brussel; email: diraerts@vub.ac.be }\ , 
Bart D'Hooghe\thanks{ Centre Leo Apostel and Department of Mathematics, Vrije Universiteit Brussel; email:   bdhooghe@vub.ac.be } \ and Mark Sioen\thanks{Department of Mathematics, Vrije Universiteit Brussel; email: msioen@vub.ac.be}}
\date{\today}
\maketitle

\begin{abstract}
\noindent The definition of  `classical state'  from \cite{A09}, used e.g. in \cite{ADDh10} to prove a decomposition theorem internally in the language of State Property Systems, presupposes as an additional datum an orthocomplementation on the property lattice of a physical system. In this paper we argue on the basis of the $(\varepsilon,d)$-model on the Poincar\'{e} sphere that a notion of \textit{topologicity} for states can be seen as an alternative (operationally foundable) classicality notion in the absence of an orthocomplementation, and compare it to the known and operationally founded concept of classicality.
\end{abstract}

\medskip

\begin{quotation}
\end{quotation}

\section{Introduction}
In earlier work we have proven that it is possible to distinguish the
`classical parts of a general State Property System', and more specifically
to derive a general decomposition theorem (Aerts \cite{A83}; Aerts,
Deses and D'Hooghe  \cite{ADDh10}). This decomposition theorem
splits off the part of the State Property System consisting of the classical
properties and classical states. 
A classical property $a$ is a property such that $a$ is actual or the orthocomplemented property $a^\perp$ is actual for any state of the State Property System. This means that the classical part that is
split off by means of the decomposition theorems in Aerts \cite{A83} and Aerts et al.
  \cite{ADDh10} is  a `deterministic part'.
Indeed, for such a classical property $a$ we can consider a test $\alpha$ such that the inverse test $\tilde\alpha$ is a test of the orthocomplemented property $a^\perp$, and then  
$\alpha$ gives with certainty the answer `yes' or gives with certainty the
answer `no' for any state of the considered physical system. Hence there is no uncertainty for such a test $\alpha$.

Often it is claimed about classical mechanics that it is a deterministic theory. The situation is however more complex, as we have observed in Aerts, Coecke, Durt
and Valckenborgh \cite{ACDV97}, more specifically in section 9 of this article,
where we have analyzed in which way we can regain the standard topology on the
sphere in the case of our considered model of a quantum spin-${\frac{1 }{2}}$-system on the Poincar\'e or Bloch sphere. The situation is indeed more complex in case one considers classical mechanics `plus' its concrete role as a theory to describe real macroscopical physical systems in the real world. Such real systems can be in states where it is difficult to give a meaning to the notion of `determined value of a physical quantity'. What we mean are the states of unstable equilibrium. Of course, formally, like their name indicates, states of unstable equilibrium are states of equilibrium, meaning that the system is in equilibrium in such states, and hence values of physical quantities have determined outcomes. In practice however, i.e. for real existing macroscopic system, such unstable equilibrium states are not really equilibrium states, since any perturbation --- in quantum
language one would say any effect of context --- will push the system outside
of such states. As a consequence a dynamics of positive feedback comes into existence, which means that the system is pushed even further away from its original state, which is the reason why the equilibrium is called
`unstable'. Of course,
classical mechanics as a formal theory `does not contain a model derived from first principles' about this positive feedback mechanism for states of unstable equilibrium. When engineers however `use' classical mechanics, the contextual effect in the case of unstable equilibrium states is certainly taken into account explicitly. If this would not be done, i.e. if these states of unstable equilibrium would be considered by engineers as states with a determined value for physical quantities, mechanical building constructions could be attempted that would collapse under the influence of a little breeze. What is interesting, and what we showed in Aerts et al. \cite{ACDV97}, is that in this model such a state of unstable equilibrium is effectively the `remaining of what in quantum mechanics is a superposition state, for the case where the fluctuations due to effects of measurements become infinitesimally small'. This means that in case we want to consider classical mechanics as a limit of quantum mechanics, the unstable equilibrium states would result as the leftover of the superposition states, the uncertainty being reduced in this specific way.

The analysis in Aerts et al. \cite{ACDV97} is made on the Poincar\'{e} sphere as a model for a quantum  spin-${1 \over 2}$-system, and we had not yet tried to generalize this finding for the case of a general system in a quantum axiomatic setting. It has remained, however,  one of the goals we had set for ourselves with respect to the problem of  capturing classical and quantum within an axiomatic approach. Indeed, in our opinion it entails a very promising approach to study the way in which classical mechanics can be considered  as a limit case of quantum mechanics. It could mean a big step ahead in the age-old `classical limit problem', a problem for which almost no progress has been made during the many years of quantum physics research. From the study of this situation in Aerts et al. \cite{ACDV97} we can deduce some aspects that give us hints of how to approach the problem in a general way. (i) Most importantly is that the standard topology of the sphere is retrieved in Aerts et al. \cite{ACDV97} for this `classical limit that consists in expressing that the fluctuations
due to measurements can be made as small as one wants'. (ii) We know that 
State Property Systems are categorically equivalent to  closure
spaces (Aerts, Colebunders, Van der Voorde and Van Steirteghem \cite{ACVdvVs99}). (iii) Additionally it is the linear closure of a Hilbert space what makes a Hilbert space State Property System a closure which is strictly not a topology. Taking into account (i), (ii) and (iii), we believe that if we introduce the requirement `that the closure which is equivalent to the State Property System is a topology', that this is a way  to express the type of classicality which we have observed and studied in Aerts et al. \cite{ACDV97} for the specific case of the Poincar\'{e} sphere. Concretely this means that we demand that for two properties $a$ and 
$b$ also the set union of the set of states contained in $a$ and the set of
states contained in $b$ corresponds to a property, let us denote it $a \cup b$.
And we introduce this requirement for any finite collection of properties, hence asking that for $(a_i)_{i=1}^n$ also $\cup_{i=1}^na_i$ is a property. We will call properties satisfying this requirement `topological properties'. In the present article we want to investigate whether it is possible to split off in a way similar than what we have done for classical properties a part of a general State Property System for such topological properties. We also want to make this investigation for the case of a State Property System which is not necessarily orthocomplemented. Indeed, for the decomposition theorem we proved in earlier work (Aerts \cite{A83} and Aerts et al. \cite{ADDh10}), an orthocomplementation of the State Property System was necessary. Parallel to our motivation for finding a more general way of approaching the problem of the classical limit, we want to investigate the role of the axiom of orthocomplementation with respect to the possibility of a decomposition theorem.

\section{State Property Systems and  Closure  Space}

\begin{definition}(\cite{A83,A09})
A  \emph{ State Property System} is a triple $(\Sigma,{\cal L},\xi)$ with $\Sigma$  a pre-ordered set (the states of the physical system), ${\cal L}$ a complete lattice (the properties of the physical system; we denote its top and bottom by $1$, resp. $0$) and $\xi : \Sigma \longrightarrow
{\cal P}({\cal L})$ satisfying the following conditions for all $p,q \in \Sigma$, all $a,b \in {\cal L}$ and $(a_i)_{i \in I}$ an arbitrary  family in ${\cal L}$ with $I$ a non-empty set:
\begin{enumerate}
\item $ 1 \in \xi(p), 0 \not\!\!{\in} \xi(p)$,
\item $p \leq q \Leftrightarrow \xi(q) \subseteq \xi(p)$,
\item $a \leq b \Leftrightarrow (\forall r \in \Sigma: a \in \xi(r) \Rightarrow b \in \xi(r))$,
\item $\bigwedge_{i \in I} a_i \in \xi(p) \Leftrightarrow \forall i \in I: a_i \in \xi(p)$.
\end{enumerate}
The statement
$a \in \xi(p)$ is interpreted as the property $a$ being actual if the considered physical system is in state $p$.
\end{definition} 
For a given State Property System $(\Sigma,{\cal L},\xi)$, the map
$$
\kappa : {\cal L} \longrightarrow {\cal P}(\Sigma) : a \mapsto 
\{p \in \Sigma \mid a \in \xi(p) \}
$$
is called the \emph{Cartan map}.
We note in passing that, with an appropriate notion of morphisms, the category {\bf Sps}
of State Property Systems is naturally equivalent to the category {\bf Cls} of 
 closure spaces (\textit{where the closure is not necessarily additive} like e.g. the closed linear span in a Hilbert space) and continuous maps (see Aerts et al. \cite {ACVdvVs99}).

Within the Geneva-Brussels approach, State Property Systems are used  for the study of physical systems.
An important feature of this approach is that it can be \emph{operationally founded} by making  use of
  \emph{tests = `yes'-`no'  experiments}.
It often is assumed that ${\cal L}$, moreover, admits an \emph{orthocomplementation}, i.e. a map
$$\cdot ' : {\cal L} \longrightarrow {\cal L}$$
which satisfies
$$(a')' = a, \;\; a \leq b \Rightarrow b' \leq a'\;\;, a \wedge a' = 0 \;\;
\textrm{and} \;\; a \vee a' = 1$$
for all $a,b \in {\cal L}$.  
Also recall from Aerts \cite{A09} that an operational justification for the existence of an orthocomplementation  on the property lattice  can be given by using the  notion  of {\em inverse} to a  given  test and the notion  of {\em ortho-test}. (One can show that ${\cal L}$ admits an orthocomplementation if every property can be tested by an ortho-test).
In fact, one can argue that orthocomplementation is one of the key ingredients in most of the mathematical axiomatic approaches to quantum mechanics (see  e.g.  Aerts and Pulmannov\'{a} \cite{AP06}, Beltrametti and Casinelli \cite{BC81}, Birkhoff and von Neumann \cite{BVn36}, Birkhoff \cite{B67}, Dorfer,Dvure\v{c}enskij and L\"{a}nger \cite{DDL96}, Dvure\v{c}enski and Pulmannov\'{a} \cite{DP00}, Foulis and Randal \cite{FR74}, Greechie \cite{G68}, Harding \cite{H09}, Jauch \cite{J68}, Kalmbach \cite{K83}, Mackey \cite{M63}, Piron \cite{P64,P76}, Pt\'{a}k and Pulmannov\'{a} \cite{PP91} and Sol\`{e}r \cite{S95}). 
The notion of classicality in this setting is defined by:

\begin{definition}(\cite{A09})
Let $(\Sigma,{\cal L},\xi) $ be a State Property System,
for which ${\mathcal L}$ \textrm{admits an orthocomplementation} $\cdot '$. Then a property $a \in {\cal L}$ is called a \emph{classical property} if
$\kappa (a) \cup \kappa(a') = \Sigma$.
\end{definition}
We recall from \cite{A09} that the notion of being a classical property can be given an operational underpinning as follows:
call a test a \emph{classical test} if in any state, either this test or its inverse is true (= gives the outcome `yes' with certainty); the  properties of the  property lattice (\textit{which is constructed out of  operational data, namely the tests}) that can be tested by a product test of a family of classical tests are precisely the classical properties.

If $(\Sigma, {\cal L},\xi)$ is a 
State Property System
for which ${\mathcal L}$ admits an orthocomplementation, we
write ${\cal C}$ for the set of all classical properties and for every $p \in
\Sigma$,
$$ \omega(p) := \bigwedge_{a \in \xi(p) \cap {\cal C} } a
$$
is called the  \emph{classical state}  of the considered physical system
if it is in state $p$. The set of all classical states is denoted $\Omega$ and it is shown in \cite{A09} that (under the conditions of Theorem 1  below) it can be augmented to a State Property System
$(\Omega,{\cal C},\xi_c)$.
  
Within this approach one can effectively model  e.g.  \emph{quantum systems with superselection rules} as is shown in the following decomposition theorem:

\begin{theorem} (\cite{A09})
If $(\Sigma,{\cal L},\xi) $ is a State Property System,
for which ${\mathcal L}$ \textrm{admits an orthocomplementation}, then
$$ (\Sigma,{\cal L},\xi) \simeq \bigoplus_{\omega \in \Omega}
(\Sigma_\omega,{\cal L}_\omega,\xi_\omega)
$$
where all the summands are \textit{totally non-classical } (i.e. top and bottom are the only classical properties).
\end{theorem}
It is important to note, however, that the notion of a classical property is only present on the level of the State Property System in the presence of an orthocomplementation on the property lattice!

Note that in absence of an orthocomplementation on ${\cal L}$, one still can consider the set ${\cal C}_{op}$ of \emph{operationally classical properties}, i.e. properties which can be tested by a test such that in any state either this test or its inverse is true,  and along with it, \emph{ operationally classical states}
$$ \omega_{op}(p) := \bigwedge_{a \in \xi(p) \cap {\cal C}_{op} } a
$$
for all $p \in \Sigma$. One  can now prove an operationally founded version of the Theorem 1 (see [2]).

\section{A motivating example}

We consider the $(\varepsilon,d)$-model on the Poincar\'{e} sphere $S^2 \subseteq 
{\mathbb  R}^3$ which was proposed in \cite{ACDV97,AD94} to study axiomatically `in between quantum and classical' situations and the classical limit, and which is an elaboration of the spin models developed in Aerts \cite{A86}, and also studied in Czachor \cite{C92} and Garola, Pykacz and Sozzo \cite{GPS06}.

\begin{figure}[htbp] %  figure placement: here, top, bottom, or page
   \centering
   \includegraphics[width=2.5in]{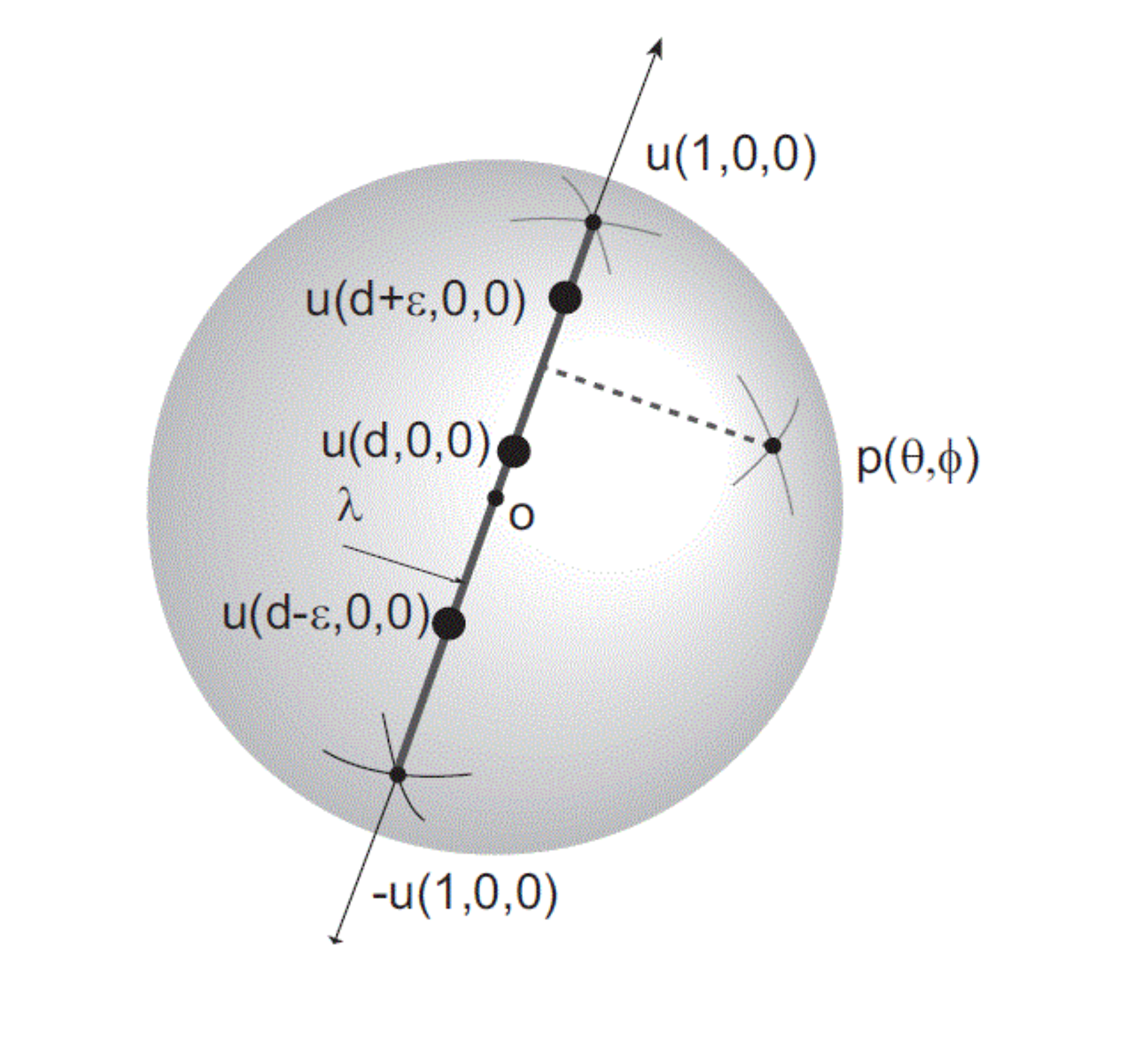} 
   \caption{A `hidden measurement model' on the Poincar\'{e} sphere}
   \label{}
\end{figure}
Assume  a particle at point $p \in S^2$ (this is the state of the system). For a vector $u \in S^2$,  $\varepsilon \in   [0,1]$ and $d \in [-1+\varepsilon,1-\varepsilon]$, consider the following test $\alpha(u,\varepsilon,d)$. Attach an elastic of length $2 \varepsilon$  centered at coordinate $d
$ along the $[-u,u]$-axis. 
Now orthogonally project $p$ onto the elastic. 
 Then   randomly cut the elastic at a point $\lambda$, where the stochastic variable 
$\lambda$ is uniformly distributed over the interval $[d -\varepsilon,
d+\varepsilon]$. If $p$ is pulled towards $u$, the outcome of $ \alpha(u,\varepsilon,d)$ is set to $\uparrow$; if $p$ is pulled towards $-u$, the outcome of $\alpha(u,\varepsilon,d)$ is set to $\downarrow$. For every $\varepsilon \in \,] 0,1]$, let
${\cal L}_\varepsilon$ be the collection of all the eigen-closed subsets of $S^2$ for the experiments $\alpha(u,\varepsilon, d)$ (all $u,d)$,  ordered by inclusion.  Consider    the State Property System
 $S_\varepsilon := (S^2,{\cal L}_\varepsilon, \xi_\varepsilon)$, with $\xi_\varepsilon(p) = \{a \in {\mathcal L}_\varepsilon \mid p \in a\}$ (all $p \in S^2$) . 
Note that the eigensets (=  sets of all  states for which  the experiment yields a certain given outcome) corresponding to the outcome $\uparrow$ (resp. $\downarrow$) for $\alpha(u,\varepsilon,d)$ are the spherical caps containing $u$ (resp. $-u$) cut off from $S^2$ by the plane  perpendicular to the $[-u,u]$-axis at coordinate $d + \varepsilon$ (resp. $d-\varepsilon$).

It is shown in \cite{ACDV97}  that for the \emph{classical limit}, i.e. taking the limit for $\varepsilon \rightarrow 0$ 
(which corresponds to taking  the categorical limit
$\varprojlim S_\varepsilon$ 
 in \textbf{Sps}, or the categorical limit of the associated eigen-closures in ${\bf Cls}$ ), one recovers the Euclidean  
\emph{topology} on the Poincar\'{e} sphere. This limit corresponds to the `zero-fluctuation' case for the measurement interaction in the sense of the Hidden Measurement Formalism for quantum mechanics  \cite{AA97} and therefore rightfully can be seen as an example of a classical deterministic system. 

This naturally leads us to proposing
topologicity = additivity, i.e.
$$\forall A,B \subseteq \Sigma: c(A \cup B) = c(A) \cup
c(B),$$  
of the closure operator $c$ corresponding to the State Property System (in the aforementioned natural equivalence ${\bf Sps} \simeq  {\bf Cls}$) as a notion of \emph{classicality} of a system. It is important to note that this notion, which we will call T-classicality,  \emph{does not assume an orthocomplementation on the property lattice}. In D'Hooghe \cite{Dh10} explicit examples are given of physical systems for which the associated property lattice of eigenclosed sets of states \emph{does not allow for any orthocomplementation}. This underlines the necesssity of defining a suitable concept of classicality which does not presuppose an orthocomplementation on the property  lattice  of a State Property System.
Apart from introducing such a notion, we also try to give an operational foundation  for it  and investigate how  it compares with the usual one in case an orthocomplementation is given.

\section{Topological states and T-classical systems}

The next proposition shows that in general the fact whether or not the property lattice of a system admits an orthocomplementation is  {\em  not preserved} when  `taking the classical limit'  in ${\bf Sps}$  like in the example of the Poincar\'{e} sphere model described in the previous paragraph. This provides another motivation for defining classicality independently of the notion of orthocomplementation.  In fact, we show that for the  lattice of closed subsets of a topological space, demanding the existence of an orthocomplementation is rather restrictive.

\begin{proposition}
Let $(X,{\cal O}_X)$ be a topological space and denote the collection of closed subsets of $X$ by  ${\mathcal C}_X$. Then the following assertions are equivalent:
\begin{enumerate}
\item  $({\cal C}_X,\subseteq)$ allows an orthocomplementation,
\item   $({\cal C}_X,\subseteq)$ is a Boolean algebra,
\item    ${\cal C}_X = {\cal O}_X$.
\end{enumerate}
\end{proposition}
{\bf Proof:}
We write $c_X$ for the closure operator associated with $(X,{\cal O}_X)$. For $A \in {\cal O}_X$,
$A^* := \bigcup \{ B \in {\cal O}_X \mid A \cap B = \emptyset  \}$ is called the  pseudo-complement   of $A$
in $({\cal O}_X,\subseteq)$ and it is easy to see that $A^* = X \setminus c_X(A)$. Since
$({\cal O}_X,\subseteq)$ is a distributive lattice, the complement of $A$ is unique if it exists and in that case has to be equal to $A^*$. One then easily obtains that $A$ is complemented in 
$({\cal O}_X,\subseteq)$ if and only if  $c_X(A) \cap c_X(X \setminus A) = \emptyset$ or, equivalently,
if and only if $A \in {\cal O}_X \cap {\cal C}_X$. In the same way, $C \in {\cal C}_X$ is complemented in 
$({\cal C}_X,\subseteq)$ if and only if $X \setminus C$ is complemented in $({\cal O}_X,\subseteq)$, if and only if $C \in {\cal O}_X \cap {\cal C}_X$. The implication $3. \Rightarrow 1.$ is obvious because if $3.$ holds, $\cdot' :{\cal C}_X \rightarrow {\cal C}_X : C \mapsto X \setminus C$  defines an orthocomplementation. Also $1. \Rightarrow 2.$ is clear since  if $1.$ holds, $({\cal C}_X , \subseteq )$
is a complete (hence bounded) distributive lattice in which every element is complemented, i.e. a Boolean algebra. Finally, assume $2.$ holds. By the  preceding  remark this implies that ${\cal C}_X \subseteq {\cal O}_X$, or equivalently, ${\cal C}_X = {\cal O}_X$.
{\hfill $\Box$}

\medskip
Note that  the condition ${\cal C}_X = {\cal O}_X$ does not imply that ${\cal C}_X = {\cal O}_X = 2^X$, the topological space $({\mathbb R},\{\emptyset,{\mathbb R}^+,{\mathbb R}_0^-,{\mathbb R}\})$ being an easy counterexample. It is also strictly stronger than $(X,{\cal O}_X)$ being zero-dimensional (i.e. that ${\cal O}_X$ is $\bigcup$-generated by a collection of open + closed subsets) as is  demonstrated  by the topological space on ${\mathbb R}$ where the topology is $\bigcup$-generated by $\{ \emptyset,{\mathbb R} \} \cup
\{[a,b[ \mid  a, b \in {\mathbb R}, a < b \}$.

\begin{definition}
Let $(\Sigma,{\cal L},\xi)$ be a State Property System. We call a property
$a \in {\cal L}$ a \emph{topological property} if
$$ \forall b \in {\cal L}: \kappa(a \vee b) \subseteq \kappa(a) \cup \kappa(b)$$
or equivalently, if 
$$ \forall b \in {\cal L}: \kappa(a \vee b) = \kappa(a) \cup \kappa(b).$$
We write ${\cal T}$ for the set of all topological properties.
For every $p \in \Sigma$,
$$\tau(p) := \bigwedge_{a \in \xi(p) \cap {\cal T }} a$$
is called the \emph{topological state} of the considered physical system if it is in state $p$, and we write $T$ for the set of all topological states.
The map 
$$\kappa_t :{\mathcal T} \longrightarrow {\mathcal P}(T) : a \mapsto \{ \tau(p) \mid p \in \kappa(a) \} $$
is called the  \emph{topological Cartan map}.
\end{definition}

\begin{proposition}
Let $(\Sigma,{\cal L},\xi)$ be a State Property System. Then for every state $p \in \Sigma$ we have that
$$\tau(p) = \bigvee_{q \in \kappa(\tau(p))} \tau(q). $$
If, moreover, the condition
${\cal T} \subseteq {\cal C}_{\textrm{op}}$ holds, then also
\begin{eqnarray}
  \tau(p) = \bigvee_{q \in \kappa(\tau(p))} \omega_{op}(q) \\
   \kappa(\tau(p)) = \bigcup_{q \in \kappa(\tau(p))} \kappa(\omega_{op}(q)).
\end{eqnarray}
\end{proposition}
{\bf Proof:} By definition of $\tau(p)$, the following implication holds:
$$ q \in \kappa(\tau(p)) \Rightarrow (\forall a \in {\cal T}: p \in \kappa(a) \Rightarrow q \in \kappa(a)),$$
hence $\tau(q) \leq \tau(p)$ for all $q \in \kappa(\tau(p))$.  Therefore  we have that
$$ \bigvee_{q \in \kappa(\tau(p))} \tau(q) \leq \tau(p) $$
and since $p \in \kappa(\tau(p))$, we have equality.
If, moreover, ${\cal T} \subseteq {\cal C}_{op}$, we have that
$$\omega_{op}(q) = \bigwedge_{a \in \xi(q) \cap {\cal C}_{op}} a \leq \tau(p) $$
for all $q \in \kappa(\tau(p))$, so
$$ \bigvee_{q \in \kappa(\tau(p))} \omega_{op}(q)   \leq \tau(p) \; \; \text{and} \;\; \bigcup_{q \in\kappa(\tau(p))} \kappa(\omega_{op}(q)) \subseteq
\kappa(\tau(p))$$
and the last inclusion  becomes  an equality because $q \in \kappa(\omega_{op}(q))$ for all $q$.
It therefore follows that
$$ \kappa(\tau (p)) = \bigcup_{q \in\kappa(\tau(p))} \kappa(\omega_{op}(q)) \subseteq
 \kappa \left(  \bigvee_{q \in \kappa(\tau(p))} \omega_{op}(q) \right)  \subseteq 
 \kappa(\tau(p))
,  $$  and because state property systems  satisfy  property determination,
$$   \bigvee_{q \in \kappa(\tau(p))} \omega_{op}(q)  = \tau(p).$$  {\hfill  $\Box$}

\medskip
Concerning an \emph{ operational foundation of the concept  topological test}, we have the following.
We call a  test $\alpha$   a \emph{topological test} if it satisfies the following condition:
$$\forall \beta: (\forall \gamma :(\alpha \leq \gamma \;\;\textrm{and}\;\; \beta \leq \gamma) \Rightarrow \gamma \; \textrm{true}) \Rightarrow
( \alpha \; \textrm{true}\; \; \textrm{or} \;\;   \beta \; \textrm{true}).$$
Then
 the
  properties of the \textit{operationally constructed} property lattice   that can be tested by a product test of a family of topological tests, are exactly the topological properties.This can be checked (elaborately but in a straightforward way) by going through the  construction of the property lattice ${\cal L}$ from the equivalence classes of tests performed in great detail 
in \cite{A09}. The line of reasoning is  analogous to the proof of the similar fact concerning classical properties being those testable by a product test of a family of classical tests, which is proved in \cite{A09} and alluded to right after Definition 2 in this paper.

\begin{definition}
A State Property System  the associated closure operator of which (in the correspondence ${\bf Sps} \simeq {\bf Cls}$) is additive, i.e. a topological closure, is called a  \emph{T-classical State-Property-System}.
\end{definition}

\begin{theorem}
Let 
$(\Sigma,{\cal L},\xi)$ be a State Property System.
Then arbitrary meets (taken in ${\cal L}$) of topological properties are again topological and therefore  ${\cal T}$ is a complete lattice with the same order and meets as ${\cal  L}$
(but in general different joins).
 If we, moreover,  define
$$  
\xi_t : T  \longrightarrow    {\cal P}({\cal T}) : \tau(p) \mapsto \xi(p)
\cap {\cal T}.
$$
Then $(T,{\cal T},\xi_t)$ is a T-classical State Property system. 
\end{theorem}
{\bf Proof:} 
Obviously $0,1 \in {\cal T}$.
Suppose $I$ is an arbitrary  (non-empty) set and $a_i \in \mathcal{T}$ for all $i \in I$. Then we have for all $b \in {\cal L}$  and $i \in	I$ that $
\kappa(a_i \vee b)=\kappa(a_i)\cup\kappa(b)$, so $(\bigwedge_{i \in I} a_i) \vee b
\le a_k\vee b$ for all $k \in I$, and hence $\kappa((\bigwedge_{i  \in I} a_i) \vee
b)\subseteq\kappa(a_k\vee b)=\kappa(a_k)\cup\kappa(b)$ for all $k \in I$. Hence \[
\kappa\left(\left(\bigwedge_{i \in I } a_i \right) \vee b\right) \subseteq
\bigcap_{i \in I} (\kappa(a_i)\cup\kappa(b))= \left(\bigcap_{i \in I} \kappa(a_i)\right)\cup\kappa(b)=\kappa\left(
\bigwedge_{i \in I } a_i\right)\cup\kappa(b).\] This proves that $\bigwedge_{i \in I}a_i\in\mathcal{T}$.
Note that $0 \in \mathcal{T}$, since $\kappa(0\vee
b)=\kappa(b)\subseteq\emptyset\cup \kappa(b)=\kappa(0)\cup\kappa(b)$ for $
b\in{\cal L}$, and also $1 \in \mathcal{T}$, because $\kappa(1\vee
b)=\kappa(1)=\Sigma=\Sigma\cup\kappa(b)=\kappa(1)\cup\kappa(b)$ for $b \in 
{\cal L}$.

The foregoing proves that $\tau(p)\in\mathcal{T}$ for $p \in \Sigma$, and
hence $T \subseteq \mathcal{T}$.
It easily follows from the third axiom in Definition 1 that $\xi_t$ is well defined. 
Define
$$
\kappa_t: {\cal T} \longrightarrow {\cal P}(T) : a \mapsto 
\{\tau(p) \in T \mid a \in \xi_t(\tau(p)) \}.
$$

Suppose $I$ is an arbitrary  (non-empty) set and $a_i \in \mathcal{T}$ for all $i \in I$.
Then  we have $\kappa_{t}(\bigwedge_{i \in I} a_i)=\{\tau(p)
\ \vert\ p\in\kappa(\bigwedge_{i \in I} a_i)\}=\{\tau(p)\ \vert\   \forall i \in I:p\in \kappa(a_i)\
\}=\bigcap_{i
\in I } \kappa_{t}(a_i)$. We have $\kappa_{t
}(0)=\{\tau(p)\ \vert\ p\in\kappa(0)\}=\emptyset$, and $\kappa_{t
}(1)=\{\tau(p)\ \vert\ p\in\kappa(1)=\Sigma\}=T$.

This proves that $(T, \mathcal{T}, \xi_{t})$ is a state
property system. We have not yet investigated `what happens with the
supremum' in the lattice $\mathcal{T}$. Consider $a, b \in \mathcal{T}$ and $
c \in {\cal L}$. We have $\kappa((a\vee b)\vee c))=\kappa(a\vee (b\vee
c))=\kappa(a)\cup\kappa(b\vee
c)=\kappa(a)\cup\kappa(b)\cup\kappa(c)=\kappa(a\vee b)\cup\kappa(c)$ which
proves that $a\vee b \in \mathcal{T}$, and we have $\kappa(a\vee
b)=\kappa(a)\cup\kappa(b)$. From this follows that the supremum $%
\vee_{i=1}^na_i$ of a finite number of elements $a_i \in {\cal T}$ is
also an element of $\mathcal{T}$, and $\kappa(\vee_{i=1}^na_i)=\cup_{i=1}^n%
\kappa(a_i)$ by induction. Indeed, suppose that this is satisfied for $n-1$,
let us prove it for $n$. We have $\kappa(\vee_{i=1}^{n}a_i\vee
c)=\kappa(\vee_{i=1}^{n-1}a_i\vee a_n\vee
c)=\kappa(\vee_{i=1}^{n-1}a_i)\cup\kappa(a_n\vee
c)=\kappa(\vee_{i=1}^{n-1}a_i)\cup\kappa(a_n)\cup\kappa(c)=\kappa(%
\vee_{i=1}^na_i)\cup\kappa(c)$, which proves that $\vee_{i=1}^na_i\in%
\mathcal{T}$. We have $\kappa(\vee_{i=1}^na_i)=\kappa(\vee_{i=1}^{n-1}a_i%
\vee a_n)=\kappa(\vee_{i=1}^{n-1}a_i)\cup\kappa(a_n)=\cup_{i=1}^{n-1}\kappa(a_i)\cup
\kappa(a_n)=\cup_{i=1}^n \kappa(a_i)$.

However, for an infinite number of elements $a_i \in \mathcal{T}$ we do not
necessarily have this equality. The supremum of such an infinite number of
elements $a_i \in \mathcal{T}$ , which we denote $\widetilde{\bigvee}$, since it
does not necessarily coincide for an infinite number of elements with $\bigvee$
which is the supremum in ${\cal L}$, is given by 
\begin{equation}
\widetilde{\bigvee}_{i \in I} a_i=\bigwedge\{  b\in\mathcal{T} \mid \forall i \in I: a_i \leq b\}
\end{equation}
The foregoing shows that $\tilde{\vee}=\vee$ for a finite number of
elements. Consider $a, b \in \mathcal{T}$, then we have 
\begin{equation}
\kappa_{t}(a \tilde{\vee} b)=\kappa_{t}(a) \cup \kappa_t(b)
\end{equation}
and also for a finite number of elements $a_i \in \mathcal{T}$ we have 
\begin{equation}
\kappa_{t}(\tilde{\vee}_{i=1}^na_i)=\cup_{i=1}^n\kappa_{t
}(a_i)
\end{equation}
which proves that the closure corresponding to the state property system $
(T, \mathcal{T}, \xi_{t})$ is  topological.
{\hfill $\Box$}

%For a state property system $(\Sigma(\mathcal{H}), \mathcal{L(H)},\kappa)$
%where $\Sigma(\mathcal{H})$ is formed by the rays of a complex Hilbert space 
%$\mathcal{H}$, hence a standard quantum mechanical state property system, we
%have that if $a \in {\cal L}(\mathcal{H})$ such that $\kappa(a \vee
%c)=\kappa(a)\cup\kappa(c)$ for $c\in {\cal L}(\mathcal{H})$, we have $a=0$
%or $a=1$, and for $p\in \Sigma(\mathcal{H})$ we have $\tau(p)=1$. Hence $%
%\mathcal{T}(\mathcal{H})=\{0,1\}$ and $T(\mathcal{H})=\{1\}$. Hence the
%topological state property system corresponding to a Hilbert space is the
%trivial one $(\{1\},\{0,1\},\kappa_{\mathcal{T}})$, where $\kappa_{\mathcal{T%
%}}(0)=\emptyset$ and $\kappa_{\mathcal{T}}(1)=1$.

\medskip
Let us now take a look at what this means for a standard quantum mechanical State Property System 
 $(\Sigma(\mathcal{H}), \mathcal{L(H)},\xi_{\cal H})$ (i.e. ${\cal H}$ is a Hilbert space,  $\Sigma(\mathcal{H})$ is formed by the rays in
${\cal H}$, ${\cal L}({\cal H})$ is the  lattice of all closed subspaces of ${\cal H}$ and $\xi_{\cal H}$ is defined by subspace inclusion).
If $a \in {\cal L}(\mathcal{H})$ such that $\kappa(a \vee
c)=\kappa(a)\cup\kappa(c)$ for all  $c\in {\cal L}(\mathcal{H})$, we  have either   $a=0$
or $a=1$. Hence $
\mathcal{T}(\mathcal{H})=\{0,1\}$ and $T(\mathcal{H})=\{1\}$, so the topological State Property System corresponding to a Hilbert space is always the trivial one given by
$(\{*\},\{0,1\},\xi) $ with $\xi(*) = \{1\}.$

\begin{theorem}
Let
$(\Sigma,{\cal L},\xi)$ be a State Property System.
 \begin{enumerate}
\item  If ${\cal L}$ admits an orthocomplementation $\cdot ' :
 {\cal L}
 \longrightarrow {\cal L}$, then the following assertions are equivalent for every property $a \in    {\cal L}$:
\begin{enumerate}
\item $a$ is classical,
\item  $a$ is topological.
\end{enumerate}
\item  If $(\Sigma,{\cal L},\xi)$ is a State Property System for which ${\cal L}$ admits an orthocomplementation and if, moreover, ${\mathcal L}$ is atomistic, then (a) and (b) are equivalent to \\
(c) $a$ is central, i.e. $ \forall b \in {\cal L}:
b =(b   \wedge a) \vee (b \wedge a')$.  (The implication $(a) \Rightarrow (c)$ even holds without the supplementary assumption of atomisticity.)
\end{enumerate} 
\end{theorem}
{\bf Proof: } Recall that under the extra assumption of atomisticity of ${\cal L}$, the equivalence of $(a)$ and $(c)$ was proved in Aerts \cite{A09} as Proposition 33, whereas it was shown
in
 Proposition 27 of \cite{A09} that classical elements are always central, even without the assumption of atomisticity. So we only need to prove 1. Here the implication $(b) \Rightarrow (a)$ is obvious since for $a \in {\cal T}$, 
$$ \Sigma = \kappa(1) = \kappa(a \vee a') = \kappa(a) \cup \kappa(a').$$ To prove the remaining implication, assume that $a \in {\cal C}$ and $b \in {\cal L}$ arbitrary. According to Proposition 27 in \cite{A09}, we have that
$$ \kappa(a \vee b)  = \kappa((b \vee a) \wedge a) \cup \kappa((b \vee a) \wedge a') = \kappa(a) \cup \kappa((b \vee a) \wedge a').$$
But since $a \in {\cal C}$, obviously $a' \in {\cal C}$, so applying  Proposition 29  from \cite{A09} yields that
$$ (b \vee a) \wedge a' = ( b \wedge a') \vee (a \wedge a') = b \wedge a'$$
and hence
$$\kappa (a \vee b)  = \kappa(a)   \cup \kappa(b \wedge a') \subseteq \kappa( a )   \cup \kappa(b) $$
so $a$ is topological. 
{\hfill $\Box$}

\medskip
We show with a counterexample that the equivalence from 1. (where one then has to replace  `classical' by  `operationally classical') in general can fail if ${\cal L}$ does not admit an orthocomplementation. To do so, consider  the $(\varepsilon,d)$-model for $\varepsilon = d = 0$. It was shown in \cite{ACDV97}  that in this case ${\cal L}$ is the closure structure generated by (i.e. the saturation of this collection w.r.t. arbitrary intersections)
$$ \{ B(u,\pi/2) \mid u \in S^2 \} \cup \{ \emptyset, S^2\}$$
where for every $u \in S^2$,$ B(u,\pi/2)$ stands for the half-sphere with $u$ as its pole and the big circle in the plane perpendicular to the $u$-axis. Then for every $u \in S^2$:
$a_u := B(u,\pi/2)$ is operationally classical but not topological, since for $b_u := \{ -u\}$, we have that
$$a_u \vee b_u = S^2.$$

\section{Conclusions}

\begin{itemize}
\item Condition 3 in Proposition 1 is strictly stronger than $(X,{\cal O}_X)$ being zero-dimensional and there are non-trivial examples where condition 3 holds and for which
${\cal O}_X \neq {\cal P}(X)$.\emph{ This shows that demanding existence of an orthocomplementation on a T-classical state property system is very restrictive}.

\item A careful examination of the proof of Theorem 1 shows that it crucially depends upon the fact that we have an orthocomplementation in order to obtain that 
$ \{ \kappa(\omega(p)) \mid p \in 
\Sigma \} $ is a partition  of $\Sigma$. This is needed to construct the ${\bf Sps}$-isomorphism in Theorem 1. Without the presence of an orthocomplementation,
$ \{ \kappa(\tau(p)) \mid p \in 
\Sigma \} $ is not necessarily a partition  of $\Sigma$, as can be observed from Proposition 2.

\item Poincar\'{e} sphere models can be used to construct an \emph{example of a State Property System in which the
property lattice ${\cal L}$ does not allow an orthocomplementation and in which there exist operationally classical properties that are not topological. }
\end{itemize}

\section{Acknowledgments}
The authors thank the anonymous referees for their valuable comments, which helped to improve the quality of the presentation.  This work was carried out within
the projects G.0234.08 and G.0405.08 of the Research Program of the Research
Foundation--Flanders (FWO).

\small{

\end{document}
\begin{thebibliography}{99}

\bibitem{A83} Aerts, D. (1983). Classical theories and non-classical theories as a special case of a more general theory. \textit{Journal of Mathematical
Physics}, \textbf{24}, pp. 2441--2453.

\bibitem{A86} Aerts, D. (1986). A possible explanation for the probabilities of quantum mechanics. {\it Journal of Mathematical Physics}, {\bf 27}, pp. 202-210. 

\bibitem{A09} Aerts, D. (2009). Quantum axiomatics. In K. Engesser, D. Gabbay and D. Lehmann (Eds.), Handbook of Quantum Logic and Quantum Structures. Elsevier, Amsterdam. 

\bibitem{AA97} Aerts, D. and Aerts, S. (1997). The hidden measurement formalism: quantum mechanics as a consequence of fluctuations on the measurement. In M. Ferrero and A. van der Merwe (Eds.), New Developments on Fundamental Problems in Quantum Physics (pp. 1--6).  Kluwer Academic, Dordrecht. 

\bibitem{ACDV97} Aerts, D., Coecke, B., Durt, T. and Valckenborgh, F. (1997).
Quantum, classical and intermediate II: the vanishing vectorspace structure. \textit{Tatra Mountains Mathematical Publications}, \textbf{10}, pp. 241--266.

\bibitem{ACVdvVs99} Aerts, D., Colebunders, E., Van der Voorde, A. and Van Steirteghem, B. (1999). State property systems and closure spaces: a study of categorical
equivalence. \textit{International Journal of Theoretical Physics}, 
\textbf{38}, pp. 359--385.


\bibitem{ADDh10} Aerts, D., Deses, D. and D'Hooghe, B. (2010). A  decomposition theorem for state
property systems.  Archive reference and link \url{http://arxiv.org/abs/quant-ph/0503083} .

\bibitem{AD94} Aerts, D. and Durt, T. (1994). Quantum, classical and intermediate, an illustrative example. \textit{Foundations of Physics},  \textbf{24}, pp. 1353--1369. 


%\bibitem{ADhS} Aerts, D., D'Hooghe, B. and Sioen, M. (2009). Decomposition theorem for state
%property systems. Preprint. Archive reference and link
%http://uk.arxiv.org/\-abs/quant-ph/0503083.

\bibitem{AP06} Aerts, D. and Pulmannov\'{a}, S. (2006).  Representation of state property systems. {\it Journal of Mathematical Physics}, \textbf{47}, pp. 1--18.

\bibitem{BC81} Beltrametti, E. G. and Cassinelli, G. (1981). The Logic of
Quantum Mechanics. Addison-Wesley, Reading, Massachusetts.

\bibitem{BVn36} Birkhoff, G. and von Neumann, J. (1936). The logic of quantum mechanics. \textit{Annals of Mathematics} \textbf{37} (4),
pp. 823--843.

\bibitem{B67} Birkhoff, G. (1967). Lattice Theory. AMS, Providence, Rhode Island.

\bibitem{C92} Czachor, M. (1992). On classical models of Spin. \textit{Foundations of Physics Letters}, \textbf{5}(3), pp. 249--264.

\bibitem{Dh10} D'Hooghe, B. (2010).   On the orthocomplementation of State-Property-Systems of contextual systems.
{\it  International Journal of Theoretical Physics}, \textbf{49} (12), pp. 3069--3084. 

\bibitem{DDL96} Dorfer, G., Dvure\v {c}enskij, A. and L\"{a}nger, H. (1996).
Symmetric difference in orthomodular lattices, {\it Mathematica Slovaca}, 
{\bf 46}, pp. 435--444. 

\bibitem{DP00}  Dvure\v {c}enskij, A. and Pulmannov\'{a}, S. (2000). New Trends in Quantum Structures. Kluwer, Dordrecht.

\bibitem{FR74} Foulis, D. J. and Randall, C. H.  (1974). Operational statistics I. Basic concepts, \textit{Journal of Mathematical
Physics}, \textbf{13}, pp. 1667--1675.

\bibitem{GPS06} Garola, C., Pykacz, J. and Sozzo, S. (2006). Quantum Machine and Semantic Realism Approach: a Unified Model,
\textit{Foundations of Physics}, \textbf{36}(6), pp. 862--882.

\bibitem{G68} Greechie, R. J. (1968).  On the structure of orthomodular lattices satisfying the chain condition.
\textit{Journal of Combinatorial Theory}, \textbf{4}, pp. 210--218.

\bibitem{H09} Harding, J. (2009). A link between quantum logic and categorical quantum mechanics. \textit{International Journal of Theoretical Physics}, \textbf{3}, 769--802.

\bibitem{J68} Jauch, J.M. (1968). Foundations of Quantum Mechanics.  Addison-Wesley, Reading, Mass..

\bibitem{K83} Kalmbach, G. (1983). Orthomodular Lattices. Academic Press, London.

\bibitem{M63}
Mackey, G. W. (1963). The Mathematical Foundations of Quantum
Mechanics. Benjamin, New York.

\bibitem{P64} Piron, C. (1964), Axiomatique quantique. {\it Helvetica Physica Acta},  \textbf{37}, pp. 439--468.

\bibitem{P76} Piron, C. (1976), Foundations of quantum physics. W. A. Benjamin, Reading, Mass..

\bibitem{PP91} Pt\'{a}k, P., and Pulmannov\'{a}, S. (1991). Orthomodular
Structures as Quantum Logics. Kluwer, Dordrecht.

\bibitem{S95} Sol\`{e}r, M.P. (1995), Characterization of Hilbert spaces by orthomodular spaces.  \textit{Communications in Algebra},
\textbf{23}, pp.  219--243.

\end{thebibliography}
